\documentclass[twocolumn,printnumbers,amsmath,amssymb,prl,verbatim]{revtex4}
\usepackage{graphicx}
\usepackage{color}

\begin{document}

\title{Connecting glass-forming ability of binary mixtures of soft particles to equilibrium melting temperatures}

\author{Yunhuan Nie, Jun Liu, Jialing Guo \& Ning Xu$^*$}

\affiliation{Hefei National Laboratory for Physical Sciences at the Microscale, CAS Key Laboratory of Microscale Magnetic Resonance and Department of Physics, University of Science and Technology of China, Hefei 230026, People's Republic of China\\
$^*$Correspondence and requests for materials should be addressed to N.X. (email: ningxu@ustc.edu.cn)}

\begin{abstract}

The glass-forming ability is an important material property for manufacturing glasses and understanding the long-standing glass transition problem. Because of the nonequilibrium nature, it is difficult to develop the theory for it. Here we report that the glass-forming ability of binary mixtures of soft particles is related to the equilibrium melting temperatures. Due to the distinction in particle size or stiffness, the two components in a mixture effectively feel different melting temperatures, leading to a melting temperature gap. By varying the particle size, stiffness, and composition over a wide range of pressures, we establish a comprehensive picture for the glass-forming ability, based on our finding of the direct link between the glass-forming ability and the melting temperature gap. Our study reveals and explains the pressure and interaction dependence of the glass-forming ability of model glass-formers, and suggests strategies to optimize the glass-forming ability via the manipulation of particle interactions.

\end{abstract}

\maketitle

\noindent {\bf Introduction}

\noindent In studies of glass transition and jamming transition \cite{debenedetti,berthier1,liu,ohern,torquato}, binary mixtures of particles are widely employed to avoid crystallization. If the two components mix up randomly, the particle size mismatch can frustrate the global structural order \cite{tanaka}. However, under certain circumstance, some binary mixtures may undergo phase separation or demixing during the solid formation, i.e., the same types of particles aggregate. Although undesirable in disordered systems, phase separation has attracted much attention in many fields \cite{bates, moreo, biben,jaeger, erlebacher, balazs,palacci,lipilong}.

The glass-forming ability (GFA), i.e., the capacity of a material to resist crystallization and maintain glassy, is fundamental in studies of glasses. A binary mixture prone to phase separation tends to form crystallites of the same type of particles, and hence has a poor GFA. Because glasses are diverse and out of equilibrium, it is difficult to establish a common understanding of the GFA. Moreover, phase separation is one of the multiple forms of crystallization. Under certain conditions, binary mixtures can also form complex crystalline or quasicrystalline structures \cite{laves,qc}, which complicates further the understanding of the GFA to fight against them. Among various interpretations, geometric and energetic frustrations are thought to be essential to the determination of the GFA \cite{xi,liy,fujita,bendert,wu,ninarello,tanaka_prx}.

For binary mixtures, varying the size ratio $\gamma$ of the large to small particles is an effective way to cause geometric frustration. However, it has been shown that a large $\gamma$ often promotes phase separation and thus results in poor GFAs for binary mixtures of hard particles, mainly arising from entropy effects \cite{biben,biben1,dijkstra,dijkstra1,dinsmore, imhof,frenkel}. To suppress phase separation, a moderate size ratio, $1<\gamma<2$, is usually adopted. However, it is still unclear if such ratios can prevent phase separation and maintain glassy states after extremely long equilibration. Even with these size ratios, the GFA is sensitive to the particle composition. It has been suggested that the GFA of binary mixtures is the best near the eutectic point or triple point when particle composition is varied \cite{tanaka_prx}.

In this work, we pay more attention to the energetic frustration on the GFA by studying binary mixtures of soft particles interacting via finite-range repulsions. In the zero temperature ($T=0$) and zero pressure ($p=0$) limit, such soft particles behave like hard ones \cite{xu,wangxipeng}. Taking a mixture in the hard particle limit as the reference, which has a decent GFA with certain values of particle size ratio and composition, we vary the pressure (density) and track the evolution of the GFA. When pressure increases, the potential energy plays a more important role and offers soft particles extra opportunities of phase separation. With the intervention of potential energy, the particle size ratio $\gamma$, the cause of geometric frustration, may excite and affect energetic frustration as well and hence affect the GFA. We aim at looking for some simple energetic criteria of the GFA and proposing some energy strategies to manipulate it, via the study of the effects of pressure and particle interaction.

We study two types of widely employed model glass-formers, binary mixtures of soft particles interacting via harmonic or repulsive Lennard-Jones (RLJ) repulsion. Both types of systems with a diameter ratio $\gamma=1.4$ and a 50:50 particle composition have been used in many studies of glasses. At low pressures, both systems exhibit a pressure independent GFA, equal to that of the hard particle counterparts. At high pressures, the GFA of RLJ systems still remains almost constant, but remarkable phase separation occurs in harmonic systems. Therefore, $\gamma=1.4$ does cause not only geometric but also energetic frustrations in soft particle systems. By performing analytic calculations, we find that the behaviors of the GFA can be explained by the pressure dependence of the melting temperatures of the two components, confirmed by our simulation results. This thus builds up a bridge between non-equilibrium and equilibrium quantities and suggests that the pressure dependence of the melting temperatures of constituent components can be an energetic precursor to evaluate the GFA of glass-formers. We also show that the GFA of the hard particle limit can be achieved at high pressures by a proper modulation of the particle stiffness. Combining our work and previous studies on the composition effects \cite{tanaka_prx}, we expect to achieve a much more comprehensive understanding of the GFA of binary mixtures with effects of composition, pressure, and particle interaction being all included.

\vspace{3mm}

\noindent {\bf Results}

\noindent {\bf Control variables of binary mixtures.} We denote the two types of particles in binary mixtures as $A$ and $B$ particles. The particle composition is quantified by the concentration of $B$ particles, $c_{\rm B}=N_{\rm B}/(N_{\rm A}+N_{\rm B})$,  with $N_{\rm A}$ and $N_{\rm B}$ being the numbers of $A$ and $B$ particles, respectively. There are two quantities to make the constituent $A$ and $B$ particles different. One is the particle size. When the diameter ratio $\gamma=\sigma_{\rm A} /\sigma_{\rm B}> 1$, $A$ and $B$ particles differ in size, where $\sigma_{\rm A}$ and $\sigma_{\rm B}$ are diameters of $A$ and $B$ particles, respectively. The other is the particle stiffness, characterized by the interaction energy scale $\epsilon_{\rm AA}$ and $\epsilon_{\rm BB}$ as defined in Eqs.~(\ref{eq_pot_harm}) and (\ref{eq_pot_rlj}) of the Methods section. $A$ particles are softer than $B$ particles if $\epsilon_{\rm AA} < \epsilon_{\rm BB}$. Here we manipulate the particle stiffness by letting $\epsilon_{\rm AA}=(1+\Delta)\epsilon_{\rm AB}$ and $\epsilon_{\rm BB}=(1-\Delta)\epsilon_{\rm AB}$ with $\Delta\in[-1,1]$. One of our primary goals is to study and understand the pressure dependence of the GFAs of widely employed model glass-formers with $\gamma=1.4$, $c_{\rm B}=0.5$, and $\Delta=0$ ($\epsilon_{\rm AA}=\epsilon_{\rm BB}=\epsilon_{\rm AB}$). This manipulation of particle stiffness can facilitate the analytic calculations with $\Delta$ being an independent variable, from which the solutions of $\Delta=0$ can be straightforwardly obtained.

In a rather different perspective from most of the previous studies, we are mainly concerned about the effects of pressure and particle interaction on the GFA. Here we mainly show results of $N=N_{\rm A}+N_{\rm B}=4096$ systems in two dimensions. Because the simulations are rather expensive, we have only repeated part of the simulations for larger systems and three-dimensional systems and find consistent results. In the following, we will first present results for harmonic and RLJ systems with $\gamma=1.4$, $c_{\rm B}=0.5$, and $\Delta=0$. Then we will attack the special case with $\gamma=1$ and $c_{\rm B}=0.5$. By varying $\Delta$, we are able to sort out potential energy effects without the interference of geometric frustration caused by the particle size distinction. Based on observations of $\gamma=1$, we will derive a generalized picture for $\gamma>1$, from which the results of systems with $\gamma=1.4$, $c_{\rm B}=0.5$, and $\Delta=0$ can be understood. Finally, we will show that the same picture applies to different values of $c_{\rm B}$.

\vspace{3mm}

\noindent {{\bf Characterization of glass-forming ability.}} For given values of $\gamma$, $c_{\rm B}$, $\Delta$, and $p$, we start with a liquid equilibrated at about four times of the melting temperature of $B$ particles as defined later, and then decrease the temperature by a small step $\delta T$ and relax the system for a duration $\Delta t$ by performing molecular dynamics simulations under constant temperature and pressure. The same procedure is repeated until a solid-like state at a temperature about one-tenth of the melting temperature is achieved. This leads to a quench rate $\kappa=\delta T / \Delta t$. We compare the GFAs over a wide range of pressures, which have been rarely studied before. It is tricky how to choose a reasonable quench rate to include the pressure effects. Here, we use a dimensionless quench rate $\tilde{\kappa}$ as defined and discussed in the Methods section, in purpose of giving supercooled liquids at different pressures comparable opportunities to relax their structures.

In the parameter space considered in this work, we have not observed the formation of complex crystalline structures within our simulation time window, so the crystallization mainly takes the form of phase separation \cite{tanaka_prx}. Therefore, the GFA can be well characterized by the degree of mixing of $A$ and $B$ particles against phase separation in the resultant solid-like states, which is quantified here by the parameters
\begin{equation}
\chi_{\rm A(B)}=\frac{1}{N_{\rm A(B)}}\sum_i\delta_{n_{{\rm s}i},n_i},
\end{equation}
where the sum is over all $A$ or $B$ particles, $\delta_{n_{{\rm s}i},n_i}$ is the Kronecker delta, $n_i$ is the number of nearest neighbors of particle $i$, and $n_{{\rm s}i}$ is the number of nearest neighbors which are the same type as particle $i$. We calculate $\chi_{\rm A}$ and $\chi_{\rm B}$ for $A$ and $B$ particles separately. Their values are close to $0$ when two components mix up well. With more particles being separated, the values grow up and approaches $1$ for well-separated states. Therefore, smaller values of $\chi_{\rm A}$ and $\chi_{\rm B}$ mean better GFAs.

Although $\chi_{\rm A}$ and $\chi_{\rm B}$ can characterize the GFA well in our work, we should realize that they may not work well to distinguish glasses and complex crystals with $A$ and $B$ particles being mixed \cite{laves,qc}. This is the limitation of the parameters when studying the GFA against the formation of complex crystals. In that case, one needs to select appropriate parameters to characterize the structural order to distinguish complex crystals from amorphous solids, which is out of the scope of current work.

\begin{figure*}
	\includegraphics[width=\textwidth]{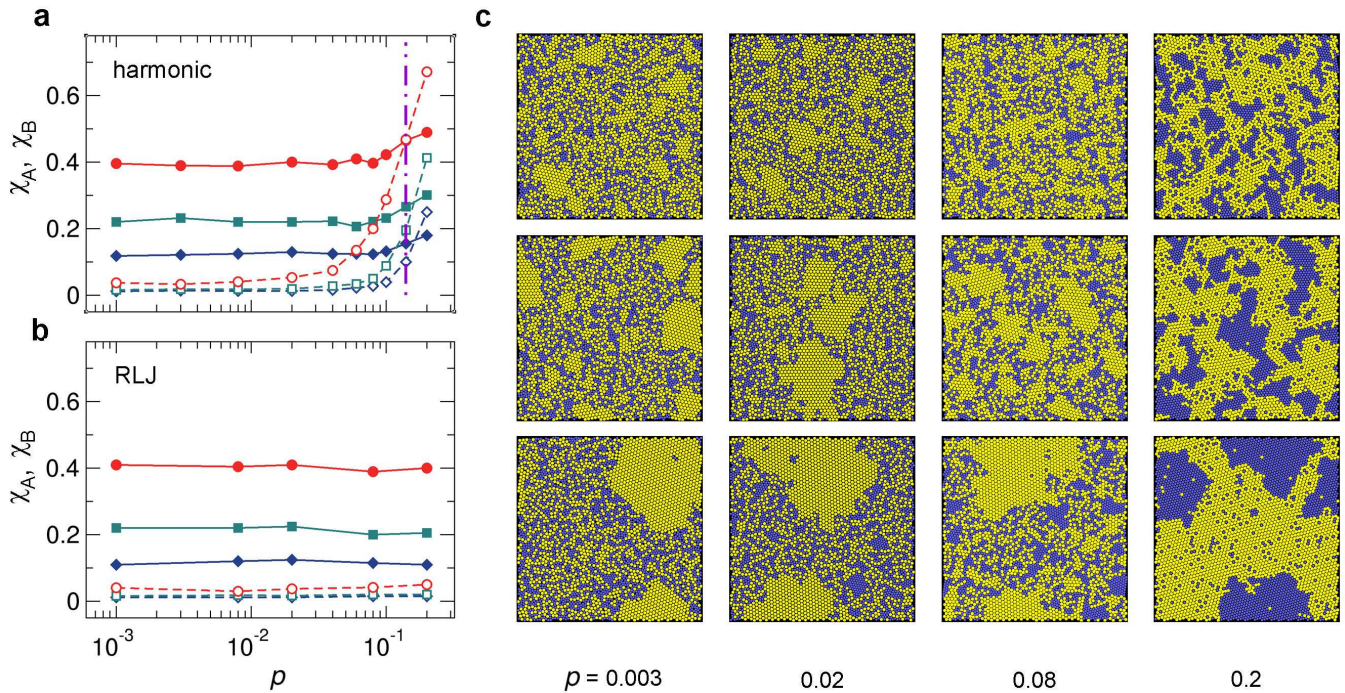}
	\caption{\label{fig:fig1}  Pressure and quench rate dependence of the glass-forming ability. {\bf a} and {\bf b} are for harmonic and RLJ systems, respectively, with $\gamma=1.4$, $c_{\rm B}=0.5$, and $\Delta=0$. The solid and empty symbols are $\chi_{\rm A}$ and $\chi_{\rm B}$, respectively.  Circles, squares, and diamonds are for dimensionless quench rate $\tilde\kappa=8.16\times{10}^{-11}$, $8.16\times{10}^{-10}$, and $8.16\times{10}^{-9}$, respectively. The lines are guides for the eye. The vertical dot-dashed line in {\bf a} shows the crossover pressure $p_{\rm n}\approx 0.14$ (also shown in Fig.~\ref{fig:fig2}a for comparison), at which the melting temperatures of $A$ and $B$ particles intersect and $\chi_{\rm A}$ is approximately equal to $\chi_{\rm B}$, as discussed in the text. {\bf c} Snapshots of solid-like states for harmonic systems. From top to bottom, $\tilde\kappa$ decreases. From left to right, pressure $p$ increases with the values being shown at the bottom. The yellow and blue disks are $A$ (large) and $B$ (small) particles. To distinguish particles, we have moderately decreased the particle diameters by $10\%\sim 25\%$.
	}
\end{figure*}

\vspace{3mm}

\noindent {\bf Glass-forming ability for $\gamma=1.4$ and $c_{\rm B}=0.5$ with $\Delta=0$.} Figures~\ref{fig:fig1}a and \ref{fig:fig1}b compare the GFAs of binary mixtures of harmonic and RLJ particles in two dimensions with $\gamma=1.4$, $c_{\rm B}=0.5$, and $\Delta=0$, so there is no cause of strong energetic frustration. These binary mixtures have been widely employed as good glass-formers in previous studies.

In the $T\rightarrow 0$ and $p\rightarrow 0$ limit when particle overlap is tiny, both harmonic and RLJ systems are equivalent to hard particle systems \cite{xu,wangxipeng} and exhibit similar GFAs to that of the hard particle counterpart with $\gamma=1.4$ and $c_{\rm B}=0.5$. As shown in Figs.~\ref{fig:fig1}a and \ref{fig:fig1}b, both $\chi_{\rm A}(p)$ and $\chi_{\rm B}(p)$ tend to approach a constant at low pressures for both systems.

Because of aging, $\chi_{\rm A}(p)$ and $\chi_{\rm B}(p)$ evolve with the quench rate $\tilde{\kappa}$. Figure~\ref{fig:fig1} compares three values of $\tilde{\kappa}$. At low pressures, $\chi_{\rm B}(p)$ (for small particles) remains small when $\tilde{\kappa}$ decreases, while $\chi_{\rm A}(p)$ (for large particles) grows up. As illustrated by snapshots in Fig.~\ref{fig:fig1}c, $A$ particles form clusters when $\tilde{\kappa}$ is small, even at low pressures. It is expected that with even smaller $\tilde\kappa$ the clustering or phase separation will be stronger. Here $A$ and $B$ particles have the same concentration. The separation can be weakened by increasing $c_{\rm B}$ to around $0.6$ \cite{tanaka_prx} in order to leave smaller rooms for $A$ particles to aggregate, as will be shown later, but it remains elusive whether the separation is inevitable as long as the waiting time is sufficiently long \cite{berthier_order,torquato1}.

The decrease of $\tilde{\kappa}$ at fixed pressure is analogous to the decrease of the compression rate at fixed temperature. The emergence of phase separation at low pressures thus suggests that with sufficiently slow compression rates binary mixtures of hard particles with $\gamma=1.4$ and $c_{\rm B}=0.5$ will undergo phase separation. Therefore, we show direct evidence suggesting that such binary mixtures widely employed as good glass-formers cannot prevent phase separation, so that thermodynamically-stable states may be phase-separated.

Figure~\ref{fig:fig1} also shows that, although behaving similarly at low pressures, harmonic and RLJ systems have significantly different GFAs at high pressures. The GFA of RLJ systems still remains almost constant in pressure. In contrast, harmonic systems undergo apparent phase separation, with $\chi_{\rm B}(p)$ increasing quickly when pressure increases. Apparently, at high pressures, some energetic frustration is excited strongly in harmonic systems, destabilizing the mixing of particles. The particle size distinction triggers such energy effects. Moreover, note that harmonic and RLJ systems have almost identical GFAs up to $p\approx 10^{-2}$, which is already far beyond the hard particle limit where particle interactions are not negligible. Then it is interesting to know why harmonic and RLJ systems can have similar GFAs beyond the hard particle limit, why RLJ systems can maintain an almost constant GFA to high pressures, and why the GFA of harmonic systems quickly drops at high pressures.

Harmonic and RLJ repulsions are distinct in their respectively bounded (soft-core) and unbounded natures. The soft-core nature of harmonic repulsion induces rich phase behaviors at high densities \cite{pamies,miller,zu,xu_cjps,miyazaki}. Unlike RLJ particles whose melting (crystallization) or glass transition temperature always increases with the increase of pressure \cite{khrapak,berthier,wanglijin,pedersen,ingebrigtsen}, harmonic particles exhibit reentrant liquid-solid transitions with the transition temperatures being non-monotonic in pressure and some related extraordinary phenomena \cite{pamies,zu,zu1,xu_cjps,wanglijin1,berthier2,zhao,schmiedeberg,likos}. We will show that it is just this non-monotonic behavior that makes harmonic systems to have dramatically distinct GFA from RLJ systems.

\vspace{3mm}

\noindent {\bf Melting temperature gap between components.} In this subsection, we will introduce two effective melting temperatures, $T_{\rm m,A}(p)$ and $T_{\rm m, B}(p)$, for $A$ and $B$ particles, respectively. We will see that they are not the actual equilibrium melting temperatures, which should be obtained from complicated calculations of the equilibrium phase diagram and would vary with $c_{\rm B}$ \cite{tanaka_prx,russo_prl}. They are instead derived from the equilibrium melting temperature of mono-component systems and the simple conversion of units. However, they turn out to work well to characterize the GFA.

An interesting feature shown in Fig.~\ref{fig:fig1}a is that $\chi_{\rm A}(p)$ and $\chi_{\rm B}(p)$ of harmonic systems intersect at roughly the same pressure $p_{\rm n}$ at different quench rates. When $p<p_{\rm n}$, $\chi_{\rm A}>\chi_{\rm B}$, so $A$ particles are easier than $B$ particles to form clusters. When $p>p_{\rm n}$, $\chi_{\rm A}<\chi_{\rm B}$. The clustering or nucleation starts below the melting temperature. If $\chi_{\rm B}>\chi_{\rm A}$, $B$ particles may experience a longer time than $A$ particles to nucleate, suggesting that the melting temperature of $B$ particles is higher, and vice versa. Therefore, we doubt whether Fig.~\ref{fig:fig1}a implies that $B$ particles have a lower (higher) melting temperature than $A$ particles when $p<p_{\rm n}$ ($p>p_{\rm n}$). If this is the case, melting temperature would play an important role in the determination of the GFA, but the question is how $A$ and $B$ particles can feel different melting temperatures in the same mixture.

\begin{figure}
	\includegraphics[width=0.48\textwidth]{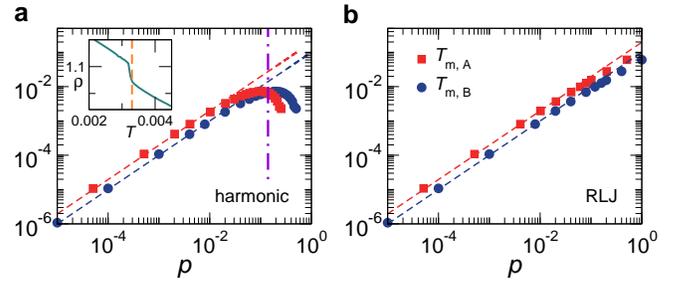}
	\caption{\label{fig:fig2} Pressure dependence of the melting temperatures. {\bf a} and {\bf b} are for harmonic and RLJ systems, respectively. Squares and circles are for $T_{\rm m,A}(p)$ and $T_{\rm m,B}(p)$, respectively, with $\gamma=1.4$ and $\Delta=0$. The dashed lines show the linear behavior. The vertical dot-dashed line in {\bf a} shows the crossover pressure at which $T_{\rm m,A}(p)$ and $T_{\rm m,B}(p)$ intersect. By comparing with Fig.~\ref{fig:fig1}a, it is roughly $p_{\rm n}$ at which $\chi_{\rm A}=\chi_{\rm B}$. The inset of {\bf a} shows how the equilibrium melting temperature $T_{\rm m}(p)$ is determined from simulation, which is also $T_{\rm m,B}(p)$ with $\Delta=0$ by definition. At a fixed pressure $p$, the melting temperature $T_{\rm m}$ (marked as the vertical dashed line) is the temperature at which the density $\rho$ undergoes an abrupt change with the decrease of temperature.
	}
\end{figure}

Note that temperature is the energy, and we are concerned about pressure effects. As defined in the Methods section, for two-dimensional mixtures, the temperature and pressure are in units of $\epsilon_{\rm AB}k_{\rm B}^{-1}$ and $\epsilon_{\rm AB}\sigma_{\rm B}^{-2}$, respectively. However, $A$ and $B$ particles also have their own temperature and pressure units, which are $\epsilon_{\rm AA}k_{\rm B}^{-1}$ and $\epsilon_{\rm AA}\sigma_{\rm A}^{-2}$ for $A$ particles and $\epsilon_{\rm BB}k_{\rm B}^{-1}$ and $\epsilon_{\rm BB}\sigma_{\rm B}^{-2}$ for $B$ particles, respectively. Therefore, in the mixture at a pressure $p$ in units of $\epsilon_{\rm AB}\sigma_{\rm B}^{-2}$, $A$ and $B$ particles effectively feel different pressure values, $P_{\rm A}$ and $P_{\rm B}$, in their own units:
\begin{equation}
P_{\rm A}=\left( p\frac{\epsilon_{\rm AB}}{\sigma_{\rm B}^2}\right) \frac{\sigma_{\rm A}^2}{\epsilon_{\rm AA}}=p\frac{\gamma^2}{1+\Delta}, \label{eq_PA}
\end{equation}
\begin{equation}
P_{\rm B}=\left( p\frac{\epsilon_{\rm AB}}{\sigma_{\rm B}^2}\right) \frac{\sigma_{\rm B}^2}{\epsilon_{\rm BB}}=p\frac{1}{1-\Delta}. \label{eq_PB}
\end{equation}

Let us denote $T_{\rm m}(p)$ as the melting temperature of mono-component systems, i.e., when $\gamma=1$ and $\Delta=0$ so that $A$ and $B$ particles are identical. In the mixture, since $A$ and $B$ particles feel different pressures in their own units, the corresponding melting temperatures are $T_{\rm m}(P_{\rm A})$ and $T_{\rm m}(P_{\rm B})$, respectively. In the common units of $\epsilon_{\rm AB}k_{\rm B}^{-1}$, the melting temperatures that $A$ and $B$ particles feel are thus
\begin{equation}
T_{\rm m,A}(p)=\left[T_{\rm m}(P_{\rm A})\frac{\epsilon_{\rm AA}}{k_{\rm B}}\right] \frac{k_{\rm B}}{\epsilon_{\rm AB}}=T_{\rm m}\left( p\frac{\gamma^2}{1+\Delta} \right)(1+\Delta), \label{eq_TA}
\end{equation}
\begin{equation}
T_{\rm m,B}(p)=\left[T_{\rm m}(P_{\rm B})\frac{\epsilon_{\rm BB}}{k_{\rm B}}\right] \frac{k_{\rm B}}{\epsilon_{\rm AB}}=T_{\rm m}\left( p\frac{1}{1-\Delta} \right)(1-\Delta). \label{eq_TB}
\end{equation}
This leads to a melting temperature gap
\begin{eqnarray}
&&\Delta T_{\rm m}(p)=T_{\rm m,A}(p) - T_{\rm m,B}(p) \nonumber\\
&&=T_{\rm m}\left( p\frac{\gamma^2}{1+\Delta} \right)(1+\Delta)-T_{\rm m}\left( p\frac{1}{1-\Delta} \right)(1-\Delta).\nonumber\\
&&\label{eq_dTm}
\end{eqnarray}
Apparently, both $\gamma$ and $\Delta$ can affect $\Delta T_{\rm m}(p)$. Note that there is a special case: When $T_{\rm m}(p)$ is linear, i.e., $T_{\rm m}(p)=Cp$ with $C$ being the system-dependent coefficient,
\begin{equation}
\Delta T_{\rm m}(p)=C(\gamma^2-1)p,\label{eq_linear}
\end{equation}
so the melting temperature gap is no longer a function of $\Delta$. We will see later that such a linear behavior is crucial to the understanding of the pressure and interaction dependence of the GFA.

Given $T_{\rm m}(p)$, Eq.~(\ref{eq_TA}) indicates that $T_{\rm m,A}(p)$ can be obtained by multiplying the pressure and temperature of $T_{\rm m}(p)$ curve by $\gamma^{-2}(1+\Delta)$ and $1+\Delta$, respectively. Correspondingly, Eq.~(\ref{eq_TB}) shows that $T_{\rm m,B}(p)$ can be obtained by multiplying the pressure and temperature of $T_{\rm m}(p)$ curve by $1-\Delta$ and $1-\Delta$, respectively. For the cases shown in Fig.~\ref{fig:fig1} with $\gamma=1.4$ and $\Delta=0$, we have $T_{\rm m,A}(p)=T_{\rm m}(\gamma^2 p)$ and $T_{\rm m,B}(p)=T_{\rm m}(p)$. Figure~\ref{fig:fig2} shows the melting temperatures against pressure for both harmonic and RLJ systems with $\gamma=1.4$ and $\Delta=0$. In the log-log scale, $T_{\rm m,A}(p)$ is obtained by just shifting the $T_{\rm m}(p)$ curve horizontally by an amount of ${\rm log}_{10}(\gamma^{-2})=-2{\rm log}_{10}1.4$, while $T_{\rm m,B}(p)$ is the same as $T_{\rm m}(p)$. If $T_{\rm m}$ monotonically increases with $p$, $T_{\rm m,A}(p)$ is always larger than $T_{\rm m,B}(p)$, as shown in Fig.~\ref{fig:fig2}b for RLJ systems. If $T_{\rm m}(p)$ is non-monotonic, $T_{\rm m,A}(p)$ and $T_{\rm m,B}(p)$ may intersect. As shown in Fig.~\ref{fig:fig2}a, $T_{\rm m}(p)$ of harmonic systems is non-monotonic in $p$, so $T_{\rm m,A}(p)=T_{\rm m,B}(p)$ at $p\approx 0.14$. In Fig.~\ref{fig:fig1}a, we display this pressure as the vertical dot-dashed line. Interestingly, it roughly agrees with $p_{\rm n}$ at which $\chi_{\rm A}=\chi_{\rm B}$.

This agreement indicates that our hypothesis that the difference between $\chi_{\rm A}$ and $\chi_{\rm B}$, as shown in Fig.~\ref{fig:fig1}, is related to the difference between melting temperatures felt by $A$ and $B$ particles is valid. This is further supported by RLJ systems. For RLJ systems, $T_{\rm m,A}(p)$ is always larger than $T_{\rm m,B}(p)$ because $T_{\rm m}(p)$ monotonically increases, consistent with the fact that $\chi_{\rm A}(p)$ is always larger than $\chi_{\rm B}(p)$, as shown in Fig.~\ref{fig:fig1}b.

Although $T_{\rm m,A}(p)$ and $T_{\rm m,B}(p)$ are not the actual equilibrium melting temperatures, the concurrence of $T_{\rm m,A}=T_{\rm m,B}$ and $\chi_{\rm A}=\chi_{\rm B}$ at $p=p_{\rm n}$ for harmonic systems suggests that $\Delta T_{\rm m}(p)$ defined here at least qualitatively reflects the correct pressure evolution of the gap between equilibrium melting temperatures. This can also be seen from the resultant solid-like states visualized in Fig.~\ref{fig:fig1}c. For harmonic systems at $p>p_{\rm n}$, $T_{\rm m,A}<T_{\rm m,B}$ and strong phase separation occurs. From the crystallization mechanism, the liquidus line is in coexistence with the $B$-solid. When $T_{\rm m,A}<T<T_{\rm m,B}$, $B$ particles crystallize, while $A$ particles can crystallize until $T<T_{\rm m,A}$. For phase-separated states obtained from finite quench rates, this sequence of the two-step crystallization would result in purer $B$-solids than $A$-solids, i.e., isolated $A$ ($B$) particles are poor (rich) in $B$-solids ($A$-solids), as shown by the snapshots at $p=0.2$ in Fig.~\ref{fig:fig1}c. The snapshots at other pressures show the opposite behavior when $p<p_{\rm n}$ and $T_{\rm m,A}>T_{\rm m,B}$.

Seen from Fig.~\ref{fig:fig2}, another interesting and important feature is that $T_{\rm m}(p)\sim p$ at low pressures for both harmonic and RLJ systems. The melting temperature curves deviate from linear at high pressures, where harmonic and RLJ systems exhibit significantly different GFAs. The comparison between Figs.~\ref{fig:fig1} and \ref{fig:fig2} shows that the GFAs maintain almost constant at fixed dimensionless quench rate $\tilde\kappa$ roughly in the pressure regime where $T_{\rm m}(p)$ is linear. Then the question is whether this is just a coincidence or implies some underlying correlations.

\vspace{3mm}

\noindent {\bf Mixing and demixing for $\gamma=1$ and $c_{\rm B}=0.5$.} In this work, we focus on the potential energy effects on the GFA. Therefore, before understanding the results of systems with $\gamma=1.4$, $c_{\rm B}=0.5$, and $\Delta=0$, we would like to discuss first a simpler case with $\gamma=1$ and $c_{\rm B}=0.5$, so that $A$ and $B$ particles have the same size and geometric frustration induced by particle size difference is absent. Then, it would be clearer to sort out the underlying physics of the evolution from mixing to demixing of two types of particles with only energetic frustration being involved, which provides us with crucial clues to understand the $\gamma=1.4$ case. In the next subsection, we will show that the findings for $\gamma=1$ can be generalized to $\gamma>1$.

With the variation of $\Delta$, $A$ and $B$ particles have different stiffness, leading to energetic frustration affecting the mixing of particles. Because of the trivial symmetry of $\gamma=1$, now we only need to vary $\Delta$ from $0$ to $1$. When $\Delta=0$, $A$ and $B$ particles are trivially the same and should statistically mix up well. When $\Delta$ increases from $0$, the distinction in particle stiffness leads to the variation in particle overlap, in analogy to the evolution with the growth of $\gamma$ \cite{tonghua}. It is thus expected that phase separation will occur when $\Delta$ is large.

\begin{figure}
	\includegraphics[width=0.48\textwidth]{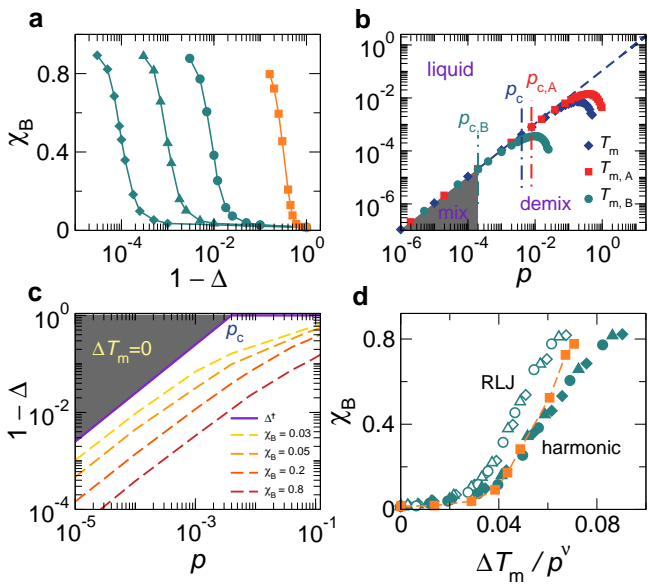}
	\caption{\label{fig:fig3} Effects of particle stiffness. Here we show results for $\gamma=1$ and $c_{\rm B}=0.5$ systems with the variation of $\Delta$ at $\tilde\kappa=1.22\times{10}^{-6}$. {\bf a} Evolution of $\chi_{\rm B}$ with $\Delta$ in solid-like states  for harmonic systems at $p=10^{-5}$ (diamonds), $10^{-4}$ (triangles), ${10}^{-3}$ (circles), and $0.12$ (squares). The lines are guides for the eye. Curves in dark green are in the pressure regime of $p<p_{\rm c}$ as defined in {\bf b}, where $T_{\rm m}(p)$ is linear. The orange curve is at $p>p_{\rm c}$ where $T_{\rm m}(p)$ is nonlinear. {\bf b} Phase diagram in the temperature and pressure plane for harmonic systems with $\Delta=0.95$. The diamonds, squares, and circles are melting temperature curves $T_{\rm m}(p)$, $T_{\rm m,A}(p)$, and $T_{\rm m,B}(p)$, respectively. The dashed line shows the linear behavior. The vertical dot-dashed lines label $p_{\rm c,B}$, $p_{\rm c}$, and $p_{\rm c,A}$, at which the melting temperature curves deviate from linear by $5\%$. The gray area below melting temperatures and to the left of $p=p_{\rm c,B}$ is where the system can mix well (labeled by `mix'). The area below melting temperatures and to the right of $p=p_{\rm c,B}$ is where $A$ and $B$ particles tend to demix (labeled by `demix'). Above melting temperatures are liquids. {\bf c} Crossover $\Delta^{\dagger}(p)$ (solid line) and the iso-$\chi_{\rm B}$ contours (dashed lines) for harmonic systems. The gray area is $\Delta\le \Delta^{\dagger}$, where $\Delta T_{\rm m}=0$. {\bf d} Scaling collapse of $\chi_{\rm B}(\Delta T_{\rm m})$ at $p=10^{-5}$ (diamonds), $10^{-4}$ (triangles), and $10^{-3}$ (circles) for harmonic (solid) and RLJ (empty) systems, with $\nu \approx 1.02$. The squares with a dashed line show the deviation from the master curve at $p=0.12>p_{\rm c}$ for harmonic systems.
	}
\end{figure}

Figure~\ref{fig:fig3}a shows examples of $\chi_{\rm B}(\Delta)$ of the resultant solid-like states at different pressures and a fixed dimensionless quench rate for harmonic systems. $\chi_{\rm A}(\Delta)$ behaves similarly to $\chi_{\rm B}(\Delta)$, so we do not show it here. In this and the next subsections, because we extend the pressure to much lower values, to use the same dimensionless quench rates $\tilde\kappa$ as in Fig.~\ref{fig:fig1} is far beyond our computational capacity. Therefore, we use faster $\tilde\kappa$. We have verified (not shown here) that the results to be presented are reproducible with different values of $\tilde\kappa$. At fixed pressure, Fig.~\ref{fig:fig3}a shows that there is a crossover $\Delta^{\rm c}$ below which $\chi_{\rm B}$ remains small and constant. When $\Delta>\Delta^{\rm c}$, $\chi_{\rm B}$ grows up, so $\Delta=\Delta^{\rm c}$ signals the onset of phase separation. Figure~\ref{fig:fig3}a also shows that $\Delta^{\rm c}$ increases when pressure decreases.

The appearing consistency between the constant GFA at low pressures and the linear $T_{\rm m}(p)$ as discussed in the previous subsection stimulates us to investigate whether the emergence of phase separation at $\Delta^{\rm c}$ is also correlated with the linearity of $T_{\rm m}(p)$. Figure~\ref{fig:fig3}b shows an example of the temperature-pressure phase diagram of harmonic systems with $\Delta=0.95$. Here we draw $T_{\rm m}(p)$ together with $T_{\rm m,A}(p)$ and $T_{\rm m,B}(p)$. Seen from Eqs.~(\ref{eq_TA}) and (\ref{eq_TB}), when $\gamma=1$, $T_{\rm m,A}(p)$ and $T_{\rm m,B}(p)$ are simply obtained by shifting the $T_{\rm m}(p)$ curve in both horizontal and vertical directions simultaneously by the same amount of ${\rm log}_{10}(1+\Delta)$ and ${\rm log}_{10}(1-\Delta)$, respectively, in the temperature-pressure plane with log-log scale. In Fig.~\ref{fig:fig3}b, we denote $p_{\rm c}$ as the crossover pressure above which $T_{\rm m}(p)$ becomes nonlinear. Here, we set $p_{\rm c}\approx 0.004$ at which $T_{\rm m}(p)$ deviates from linear by $5\%$. Correspondingly, Eqs.~(\ref{eq_TA}) and (\ref{eq_TB}) indicate that $T_{\rm m,A}(p)$ and $T_{\rm m,B}(p)$ become nonlinear at $p_{\rm c,A}=p_{\rm c}(1+\Delta)$ and $p_{\rm c,B}=p_{\rm c}(1-\Delta)$, respectively , when $\gamma=1$. Because $\Delta>0$, $T_{\rm m,A}(p)$ and $T_{\rm m,B}(p)$ collapse and are both linear when $p<p_{\rm c,B}$. Seen from Eq.~(\ref{eq_linear}), the melting temperature gap $\Delta T_{\rm m}(p)=0$ when $p<p_{\rm c,B}$, and $\Delta T_{\rm m}(p)>0$ otherwise. Therefore, in the pressure regime where $p<p_{\rm c}$, $p=p_{\rm c,B}=p_{\rm c}(1-\Delta^{\dagger})$ sets a crossover
\begin{equation}
\Delta^{\dagger}(p)=1-\frac{p}{p_{\rm c}}, \label{eq_delta}
\end{equation}
below (above) which $\Delta T_{\rm m}=0$ ($\Delta T_{\rm m}>0$). In the pressure regime where $p>p_{\rm c}$, any nonzero $\Delta$ leads to $\Delta T_{\rm m}\neq 0$, so $\Delta^{\dagger}=0$. Note that both $\Delta^{\dagger}$ and $\Delta^{\rm c}$ increase when pressure decreases. It is then natural to ask whether they are related.

Figure~\ref{fig:fig3}c shows $\Delta^{\dagger}(p)$ together with the iso-$\chi_{\rm B}$ contours for harmonic systems at fixed $\tilde\kappa$. With the decrease of $\chi_{\rm B}$, the contours approach $\Delta^{\dagger}(p)$. The contours may shift upward gradually with the decrease of $\tilde\kappa$. With current computational capacity, we expect from Fig.~\ref{fig:fig3}c that $\Delta^{\dagger}$ agrees with $\Delta^{\rm c}$. Next, we will show that this expectation is valid.

At fixed pressure, the melting temperature gap $\Delta T_{\rm m}$ is the function of $\Delta$, as shown by Eq.~(\ref{eq_dTm}). Therefore, $\chi_{\rm B}(\Delta)$ shown in Fig.~\ref{fig:fig3}a can be converted to $\chi_{\rm B}(\Delta T_{\rm m})$. This functional relation establishes the connection between the GFA characterized by $\chi_{\rm B}$ and the melting temperature gap $\Delta T_{\rm m}$ for $\gamma=1$. Figure~\ref{fig:fig3}d shows $\chi_{\rm B}(\Delta T_{\rm m})$ curves at different pressures for both harmonic and RLJ systems. At all pressures, $\chi_{\rm B}$ has a minimum when $\Delta T_{\rm m}=0$. With the increase of $\Delta T_{\rm m}$, $\chi_{\rm B}$ increases and phase separation emerges. Interestingly, in the pressure regime where $p<p_{\rm c}$ and $T_{\rm m}(p)$ is linear, all $\chi_{\rm B}(\Delta T_{\rm m})$ curves can collapse nicely onto the same master curve, when $\chi_{\rm B}$ is plotted against $\Delta T_{\rm m}/p^{\nu}$ with $\nu\approx 1.02$ for both harmonic and RLJ repulsions. Figure~\ref{fig:fig3}d also shows that the scaling collapse stops working when $p>p_{\rm c}$, highlighting the important role of the linear $T_{\rm m}(p)$.

The scaling collapse indicates that at different pressures lower than $p_{\rm c}$, in order to reach the same degree of particle mixing or demixing, the variation of particle stiffness ($\Delta$) needs to cause a melting temperature gap $\Delta T_{\rm m}$ proportional to pressure. In this pressure regime, $T_{\rm m,A}(p)$ is always linear, i.e., $T_{\rm m,A}(p)=Cp$. From Eq.~(\ref{eq_dTm}), we have
\begin{equation}
\frac{\Delta T_{\rm m}(p)}{p} = C - T_{\rm m}\left( \frac{p}{1-\Delta}\right)\frac{1-\Delta}{p}, \label{eq_dTm1}
\end{equation}
when $\Delta >\Delta^{\dagger}$ and $\Delta T_{\rm m}>0$. The right hand side of Eq.~(\ref{eq_dTm1}) is constant in pressure for a given $\chi_{\rm B}$. Because $T_{\rm m, B}(p)$ is nonlinear when $\Delta T_{\rm m}>0$ (so is $T_{\rm m}(\frac{p}{1-\Delta})$ on the right hand side of Eq.~(\ref{eq_dTm1})), $\frac{p}{1-\Delta}$ must be a constant. This leads to
\begin{equation}
\Delta(\chi_{\rm B})= 1-a(\chi_{\rm B})p, \label{eq_deltachi}
\end{equation}
where $a$ is a $\chi_{\rm B}$-dependent prefactor. Equation~(\ref{eq_deltachi}) sets the pressure dependent values of $\Delta$ to reach the same $\chi_{\rm B}$ at $p<p_{\rm c}$. Note that $\Delta^{\dagger}(p)$ in Eq.~(\ref{eq_delta}) shows exactly the same form of pressure dependence. Therefore, Eq.~(\ref{eq_delta}) is actually a direct consequence of the scaling collapse in the $\Delta T_{\rm m}\rightarrow 0$ limit. On the other hand, $\Delta^{\rm c}$ defined above follows Eq.~(\ref{eq_deltachi}) as well, corresponding to the minimum $\chi_{\rm B}$ where $\Delta T_{\rm m}=0$. Therefore, the scaling collapse naturally suggests that $\Delta^{\dagger}=\Delta^{\rm c}$. Consequently, for $\gamma=1$, phase separation can occur at a given pressure $p$ only when $\Delta>\Delta^{\dagger}$ and $\Delta T_{\rm m}>0$.

Equation~(\ref{eq_delta}) also indicates that $\Delta^{\dagger}\rightarrow 1$ when $p\rightarrow 0$, so $\epsilon_{\rm BB}\rightarrow0$ and $B$ particles cannot feel the existence of each other. In the $p\rightarrow 0$ limit, as long as $\Delta<1$ and there are still repulsions among $B$ particles, $A$ and $B$ particles become identical hard particles when $\gamma=1$ and there will be no phase separation. This is consistent with our $\Delta T_{\rm m}$ argument, because it is impossible to cause $\Delta T_{\rm m}>0$ in the $p\rightarrow 0$ limit.

\vspace{3mm}

\noindent {\bf Glass-forming ability for $\gamma>1$ and $c_{\rm B}=0.5$.} Inspired by the results of $\gamma=1$, in this subsection, we will generalize the connection between $\chi_{\rm A}$ ($\chi_{\rm B}$) and $\Delta T_{\rm m}$ to $\gamma> 1$. The results of $\gamma=1$ are thus naturally incorporated into the generalized picture. Our arguments will be examined by simulation results of $\gamma=1.4$ and $c_{\rm B}=0.5$ without loss of generality. The observations of systems with $\gamma=1.4$, $c_{\rm B}=0.5$, and $\Delta=0$ presented above will then be explained. Here we also vary $\Delta$ as done for $\gamma=1$. Because $\gamma>1$ and $A$ and $B$ particles have different sizes, $\Delta$ and $-\Delta$ apparently correspond to two different systems. We thus need to restore the range of $\Delta$ to $[-1,1]$.

\begin{figure*}
	\includegraphics[width=0.95\textwidth]{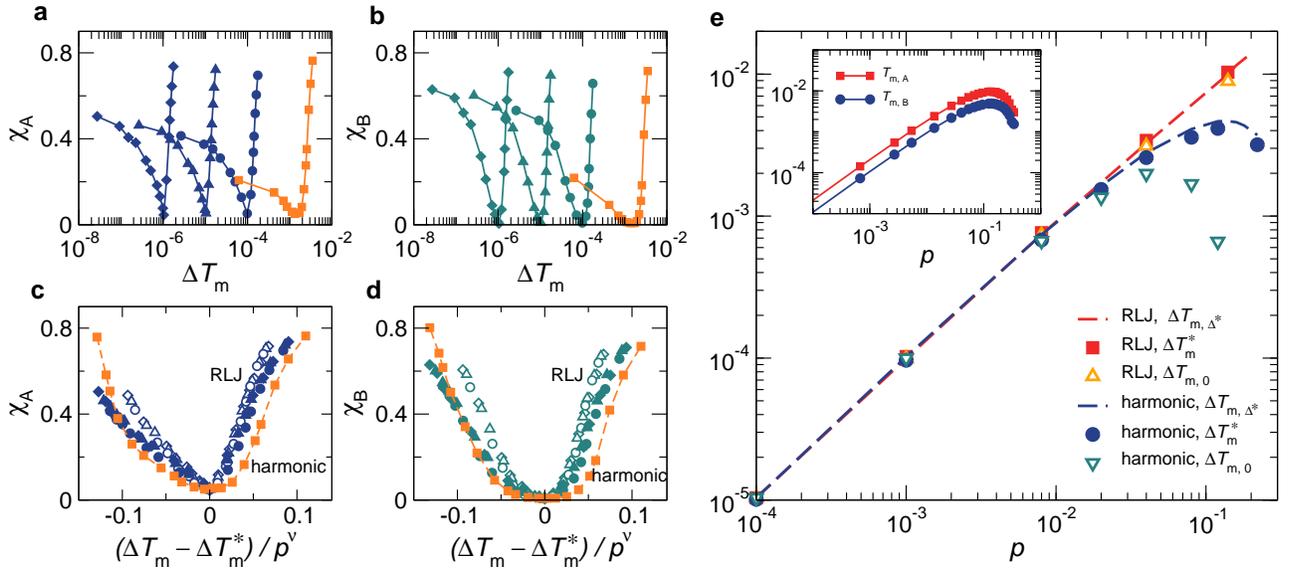}
	\caption{\label{fig:fig4} Role of the melting temperature gap. Here we show results for $\gamma=1.4$ and $c_{\rm B}=0.5$ systems with the variation of $\Delta$ at $\tilde\kappa=1.22\times{10}^{-6}$. {\bf a,b} $\chi_{\rm A}(\Delta T_{\rm m})$ and $\chi_{\rm B}(\Delta T_{\rm m})$ of solid-like states for harmonic systems at $p=10^{-5}$ (diamonds), $10^{-4}$ (triangles), $10^{-3}$ (circles), and $0.02$ (squares). The curves in orange are at $p>p_{\rm c}$ where $T_{\rm m}(p)$ is nonlinear, while other curves are at $p<p_{\rm c}$. {\bf c,d} Scaling collapse of $\chi_{\rm A}(\Delta T_{\rm m})$ and $\chi_{\rm B}(\Delta T_{\rm m})$ for harmonic (solid) and RLJ (empty) systems at $p=10^{-5}$ (diamonds), $10^{-4}$ (triangles), $10^{-3}$ (circles), with $\nu \approx 1.02$. $\Delta T_{\rm m}^*$ is the value of $\Delta T_{\rm m}$ at which $\chi_{\rm A}$ and $\chi_{\rm B}$ reach the minimum, as shown in {\bf a} and {\bf b}. The squares with a dashed line show the violation of the scaling at $p=0.02>p_{\rm c}$ for harmonic systems. {\bf e} Comparison of melting temperature gaps $\Delta T_{\rm m}^*(p)$, $\Delta T_{{\rm m},\Delta^*}(p)$ with $\Delta=\Delta^*$ defined in Eq.~(\ref{eq_deltac}), and $\Delta T_{{\rm m},0}(p)$ with $\Delta=0$ for harmonic and RLJ systems. The inset shows $T_{\rm m,A}(p)$ and $T_{\rm m,B}(p)$ with $\Delta=\Delta^*$.
	}
\end{figure*}

According to Eq.~(\ref{eq_dTm}), at fixed pressure $p$, the melting temperature gap $\Delta T_{\rm m}$ varies with $\Delta$, so we are able to establish the functional relation between the GFA and $\Delta T_{\rm m}$. Figures~\ref{fig:fig4}a and \ref{fig:fig4}b show $\chi_{\rm A}(\Delta T_{\rm m})$ and $\chi_{\rm B}(\Delta T_{\rm m})$ for harmonic systems with $\gamma=1.4$ and $c_{\rm B}=0.5$ at different pressures and a given dimensionless quench rate $\tilde\kappa$. All curves reach the minimum at $\Delta T_{\rm m}=\Delta T_{\rm m}^*$. In contrast to $\gamma=1$ for which $\Delta T_{\rm m}^*=0$, $\Delta T_{\rm m}^*$ is nonzero and varies with pressure for $\gamma=1.4$.

Analogous to $\gamma=1$, Figs.~\ref{fig:fig4}c and \ref{fig:fig4}d show that for both harmonic and RLJ systems $\chi_{\rm A}(\Delta T_{\rm m})$ and $\chi_{\rm B}(\Delta T_{\rm m})$ curves at different pressures lower than $p_{\rm c}$ can also collapse onto the same master curves, when $\chi_{\rm A}$ and $\chi_{\rm B}$ are plotted against $(\Delta T_{\rm m}-\Delta T_{\rm m}^*)/p^{\nu}$ with $\nu\approx 1.02$. Therefore, the scaling collapse shown in Fig.~\ref{fig:fig3}d for $\gamma=1$ is just a special case with $\Delta T_{\rm m}^*=0$. When $p>p_{\rm c}$, the scaling collapse breaks down.

For $\gamma=1$, we have also shown that there is a range of $\Delta$ where $\Delta T_{\rm m}=\Delta T_{\rm m}^*=0$, i.e., $-\Delta^{\dagger}<\Delta <\Delta^{\dagger}$. Then it is natural to ask whether there also exists a range of $\Delta$ within which $\Delta T_{\rm m}=\Delta T_{\rm m}^*$ for $\gamma>1$. Another interesting question is what determines $\Delta T_{\rm m}^*$.

Because $T_{\rm m}(p)$ is linear when $p<p_{\rm c}$, according to Eqs.~(\ref{eq_dTm}) and (\ref{eq_linear}), at a given $p<p_{\rm c}$, there indeed exist a range of $\Delta$ within which $\Delta T_{\rm m}$ is constant, as long as both $T_{\rm m, A}(p)$ and $T_{\rm m,B}(p)$ are still linear. As shown in Fig.~\ref{fig:fig2}, $\gamma=1.4$ leads to the separation of $T_{\rm m,A}(p)$ and $T_{\rm m,B}(p)$ when $\Delta =0$. When $\Delta$ varies from $0$, $T_{\rm m,A}(p)$ and $T_{\rm m,B}(p)$ curves in Fig.~\ref{fig:fig2} will shift in both horizontal and vertical directions simultaneously by the same amount of ${\rm log}_{10}(1+\Delta)$ and ${\rm log}_{10}(1-\Delta)$, respectively, in the log-log scale. According to Eqs.~(\ref{eq_TA}) and (\ref{eq_TB}), at a given $p<p_{\rm c}$, $T_{\rm m,A}(p)$ or $T_{\rm m,B}(p)$ become nonlinear when $\Delta$ is below
\begin{equation}
\Delta_{1}^{\dagger}=\gamma^{2}\frac{p}{p_{\rm c}}-1, \label{eq_delta1}
\end{equation}
or above
\begin{equation}
\Delta_{2}^{\dagger}=1-\frac{p}{p_{\rm c}}. \label{eq_delta2}
\end{equation}
In the $p\rightarrow 0$ limit, $\Delta_{1}^{\dagger}$ and $\Delta_{2}^{\dagger}$ approach $-1$ and $1$, respectively, consistent with our previous discussions for $\gamma=1$. At low pressures, $\Delta_2^{\dagger}$ is apparently larger than $\Delta_1^{\dagger}$. When $\Delta_1^{\dagger}\le \Delta \le \Delta_2^{\dagger}$ at a given $p$, $\Delta T_{\rm m}$ remains constant. If such a constant $\Delta T_{\rm m}$ is $\Delta T_{\rm m}^*$, the scaling collapse shown in Figs.~\ref{fig:fig4}c and \ref{fig:fig4}d also validates that $\Delta_1^{\dagger}$ and $\Delta_2^{\dagger}$ are the onsets of the growth of $\chi_{\rm A}$ and $\chi_{\rm B}$ and the weakening of the GFA, as observed for $\gamma=1$.

With the increase of pressure, $\Delta_1^{\dagger}$ and $\Delta_2^{\dagger}$ approach each other, until arriving at a critical pressure
\begin{equation}
p_{\rm e}=\frac{2}{\gamma^2+1}p_{\rm c}, \label{pe}
\end{equation}
where
\begin{equation}
\Delta_1^{\dagger}=\Delta_2^{\dagger}=\Delta^*=\frac{\gamma^2-1}{\gamma^2+1}.\label{eq_deltac}
\end{equation}
When $p>p_{\rm e}$, any variation of $\Delta$ normally causes the change of $\Delta T_{\rm m}$, so there is no interval of $\Delta$ within which $\Delta T_{\rm m}$ remains constant. Therefore, the condition $p<p_{\rm c}$ used above should be generally modified to $p<p_{\rm e}$.  When $\gamma=1$, Eqs.~(\ref{eq_delta1})-(\ref{eq_deltac}) lead to exactly the same results as shown in the previous subsection, with $\Delta_2^{\dagger}=-\Delta_1^{\dagger}=\Delta^{\dagger}$, $p_{\rm e}=p_{\rm c}$, and $\Delta^*=0$. For $\gamma=1$, $\Delta^*=0$ corresponds to the best mixing of particles with $\Delta T_{\rm m}^*=0$ at all pressures. Is it possible that $\Delta^*$ defined in Eq.~({\ref{eq_deltac}) sets $\Delta T_{\rm m}^*$ for $\gamma>1$ as well?

When plugging in $\Delta=\Delta^*$, Eqs.~(\ref{eq_TA}) and (\ref{eq_TB}) turn to
\begin{equation}
T_{\rm m,A}(p)=T_{\rm m}\left( p\frac{\gamma^2+1}{2} \right)\frac{2\gamma^2}{\gamma^2+1},
\end{equation}
\begin{equation}
T_{\rm m,B}(p)=T_{\rm m}\left( p\frac{\gamma^2+1}{2} \right)\frac{2}{\gamma^2+1}.
\end{equation}
Interestingly, $T_{\rm m,A}(p)=\gamma^2 T_{\rm m,B}(p)$, so in the log-log scale $T_{\rm m, A}(p)$ is obtained by simply shifting $T_{\rm m,B}(p)$ upward by an amount of ${\rm log}_{10}(\gamma^2)$. Then both $T_{\rm m,A}(p)$ and $T_{\rm m,B}(p)$ become nonlinear when $p>p_{\rm e}$.

In the inset of Fig.~\ref{fig:fig4}e, we show $T_{\rm m,A}(p)$ and $T_{\rm m,B}(p)$ for harmonic systems with $\gamma=1.4$ and $\Delta=\Delta^*$. In the log-log scale, they are parallel along the vertical direction. We calculate the melting temperature gap $\Delta T_{{\rm m},\Delta^*}(p)$ at $\Delta=\Delta^*$. Surprisingly, it agrees well with $\Delta T_{\rm m}^*(p)$ over the whole range of pressures studied and for both harmonic and RLJ systems, as shown in Fig.~\ref{fig:fig4}e. This agreement confirms our conjecture that there exists a pressure-independent $\Delta^*$ at which the GFA or the degree of particle mixing is the strongest, equal to that of the hard particle counterpart. When $p>p_{\rm e}$, any deviation of $\Delta$ from $\Delta^*$ causes $\Delta T_{\rm m}\neq \Delta T_{\rm m}^*$ and the weakening of the GFA. When $p<p_{\rm e}$, $\Delta^*$ lies in the interval of $\Delta\in [\Delta_1^{\dagger},\Delta_2^{\dagger}]$, within which $\Delta T_{\rm m}=\Delta T_{\rm m}^*$ and the GFA can maintain the best.

Now we are able to understand the results of systems with $\gamma=1.4$, $c_{\rm B}=0.5$, and $\Delta=0$ discussed earlier. From Eqs.~(\ref{eq_delta1}) and (\ref{eq_delta2}), we can see that $\Delta=0$ lies in $[\Delta_1^{\dagger}, \Delta_2^{\dagger}]$ when $p<\gamma^{-2}p_{\rm c}$, so the GFA can maintain the best. When $p>\gamma^{-2}p_{\rm c}$, $\Delta=0$ lies outside of $[\Delta_1^{\dagger}, \Delta_2^{\dagger}]$ and is away from $\Delta^*$ when $p>p_{\rm e}$. Therefore, $\Delta T_{{\rm m},0}$, the melting temperature gap with $\Delta=0$, deviates from $\Delta T_{\rm m}^*$ and the GFA becomes worse. How much $\Delta T_{{\rm m},0}$ deviates from $\Delta T_{\rm m}^*$ determines how worse the GFA becomes. Figure~\ref{fig:fig4}e also compares $\Delta T_{{\rm m},0}(p)$ with $\Delta T_{\rm m}^*(p)$. At low pressures, they agree very well, as expected. At high pressures, $\Delta T_{{\rm m},0}$ deviates from $\Delta T_{\rm m}^*$. For RLJ systems, the deviation is small, so the GFA is still close to the best. In contrast, for harmonic systems, the deviation significantly increases with the increase of pressure, so the GFA becomes much worse and remarkable phase separation occurs. The more essential reason of such a distinction between harmonic and RLJ systems is that harmonic systems have a much more nonlinear $T_{\rm m}(p)$.

Seen from Figs.~\ref{fig:fig4}a-\ref{fig:fig4}b, the minimum values of $\chi_{\rm A}$ and $\chi_{\rm B}$ with $\Delta=\Delta^*$ are almost constant in pressure. This explains why the GFA remains constant at low pressures for harmonic and RLJ systems with $\gamma=1.4$, $c_{\rm B}=0.5$, and $\Delta=0$. More interestingly, this suggests that the GFA of hard particle systems can be maintained to high pressures far beyond the hard particle limit, by properly modulating the particle interactions and causing the interplay between geometric and energetic frustrations. To apply $\Delta^*$ in Eq.~(\ref{eq_deltac}) is one solution for binary mixtures, which should work generally for other types of particle interactions.

\vspace{3mm}

\noindent {\bf Dependence on $c_{\rm B}$.}
In previous subsections, we have focused on $c_{\rm B}=0.5$. Figure~\ref{fig:fig5} verifies that our major findings of the pressure and particle interaction effects on the GFA hold for other values of $c_{\rm B}$. For given $c_{\rm B}$, $\gamma$, and $p$, we tune $\Delta$ and vary the melting temperature gap $\Delta T_{\rm m}$, exactly as done for $c_{\rm B}=0.5$.

\begin{figure}
	\includegraphics[width=0.48\textwidth]{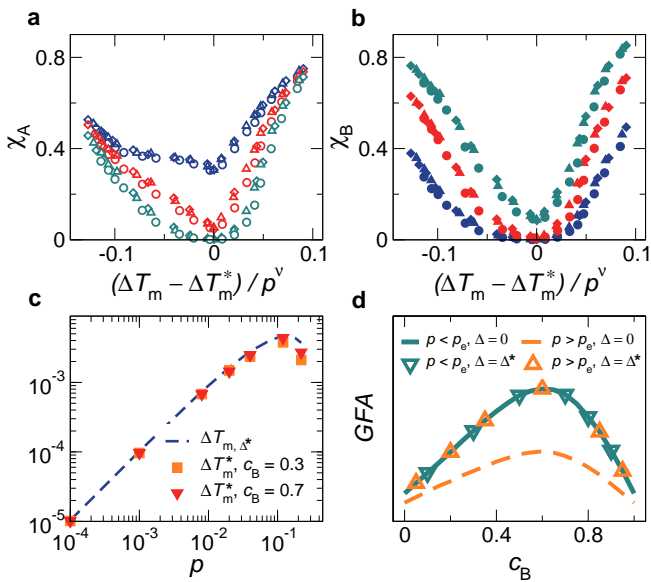}
	\caption{\label{fig:fig5} Particle composition dependence. Here we show results of harmonic systems with $\gamma=1.4$. {\bf a,b} Scaling collapse of $\chi_{\rm A}(\Delta T_{\rm m})$ and $\chi_{\rm B}(\Delta T_{\rm m})$ of solid-like states at $p=10^{-5}$ (diamonds), $10^{-4}$ (triangles), and $10^{-3}$ (circles) quenched via the rate $\tilde\kappa=1.22\times 10^{-6}$. $\Delta T_{\rm m}^*$ is the value of $\Delta T_{\rm m}$ at which $\chi_{\rm A}$ and $\chi_{\rm B}$ reach the minimum, and $\nu \approx 1.02$. Three values of $c_{\rm B}$ are presented: $0.3$ (blue), $0.5$ (red), and $0.7$ (green). {\bf c} Comparison of melting temperature gaps $\Delta T_{\rm m}^*(p)$ for $c_{\rm B}=0.3$ and $0.7$ and $\Delta T_{{\rm m},\Delta^*}(p)$ with $\Delta=\Delta^*$ defined in Eq.~(\ref{eq_deltac}). They agree well, as shown in Fig.~\ref{fig:fig4}e for $c_{\rm B}=0.5$. {\bf d} Schematic plot of the GFA against $c_{\rm B}$ on both sides of $p_{\rm e}$ defined in Eq.~(\ref{pe}) and at $\Delta=0$ and $\Delta^*$.
	}
\end{figure}

In this subsection, we take harmonic systems with $\gamma=1.4$ as the example. As shown in Figs.~\ref{fig:fig5}a and \ref{fig:fig5}b, for $c_{\rm B}=0.3$ and $0.7$, $\chi_{\rm A}$ and $\chi_{\rm B}$ reach the minimum at a pressure-dependent $\Delta T_{\rm m}^*$. Curves of $\chi_{\rm A}(\Delta T_{\rm m})$ and $\chi_{\rm B}(\Delta T_{\rm m})$ at different pressures lower than $p_{\rm e}$ defined in Eq.~(\ref{pe}) can collapse well when plotted against $(\Delta T_{\rm m}-\Delta T_{\rm m}^*)/p^{\nu}$ with $\nu\approx 1.02$. Therefore, as for $c_{\rm B}=0.5$, the GFA of the hard particle limit can be sustained with the increase of pressure, as long as $\Delta T_{\rm m}(p)$ remains linear.

Figure~\ref{fig:fig5}c reproduces the agreement between $\Delta T_{\rm m}^*$ and $\Delta T_{\rm m, \Delta^*}$ at $\Delta=\Delta^*$ for $c_{\rm B}=0.3$ and $0.7$, as shown in Fig.~\ref{fig:fig4}e for $c_{\rm B}=0.5$. For all values of $c_{\rm B}$ studied, $\Delta=0$ leads to a $\Delta T_{\rm m}$ equal to $\Delta T_{\rm m}^*$ when $p<p_{\rm e}$, as already discussed for $c_{\rm B}=0.5$ in the previous subsection. Therefore, $\Delta=0$ can maintain the GFA of the hard particle limit until $p=p_{\rm e}$, above which the GFA becomes weaker. However, when $\Delta=\Delta^*$, Figs.~\ref{fig:fig4}e and \ref{fig:fig5}c show that $\Delta T_{\rm m, \Delta^*}\approx \Delta T_{\rm m}^*$, so the GFA of the hard particle limit can be maintained over a wide range of pressures and beyond $p=p_{\rm e}$.

Seen from Figs.~\ref{fig:fig5}a and \ref{fig:fig5}b, a remarkable change with $c_{\rm B}$ is the variation of the minimum values of $\chi_{\rm A}$ and $\chi_{\rm B}$, reflecting the variation of the GFA of the hard particle limit with $c_{\rm B}$. The GFA can be approximately quantified by $\chi=\chi_{\rm A}(1-c_{\rm B})+\chi_{\rm B}c_{\rm B}$. Figures~\ref{fig:fig5}a and \ref{fig:fig5}b suggest that the GFA of $c_{\rm B}=0.3$ is apparently weaker than those of $c_{\rm B}=0.5$ and $0.7$, while the GFAs of $c_{\rm B}=0.5$ and $0.7$ are similar with that of $c_{\rm B}=0.5$ seeming a little bit stronger. The GFAs in the $c_{\rm B}=0$ and $1$ limits are trivially the weakest. We also find (not shown here) that the best GFA associated with the minimum $\chi$ occurs between $c_{\rm B}=0.5$ and $0.7$.

The evolution of the GFA with the increase of $c_{\rm B}$ characterized in our work is consistent with previous results of binary mixtures of hard disks with $\gamma=1.4$ from the calculations of the interface energy \cite{tanaka_prx}. Based on our results, we can now qualitatively extend the $c_{\rm B}$ dependence of the GFA to various pressures. In Fig.~\ref{fig:fig5}d, we schematically plot the GFA against $c_{\rm B}$ for different pressures and for $\Delta=0$ and $\Delta^*$, respectively. For $\Delta=0$, it can be expected that the GFA($c_{\rm B}$) curve of the hard particle limit can be maintained until $p=p_{\rm e}$. When $p>p_{\rm e}$ and the GFA becomes weaker, the whole GFA curve will shift down with the increase of pressure. For $\Delta=\Delta^*$, we would expect that the GFA($c_{\rm B}$) curve of the hard particle limit can be maintained beyond $p=p_{\rm e}$.

In the perspective of the empirical argument of the connection between the GFA and eutectic point, the GFA can be evaluated from the depression of the equilibrium melting temperature with the variation of $c_{\rm B}$ \cite{tanaka_prx,tanaka}. However, the argument may not be easily applied to compare the GFAs at different pressures or densities, purely from the comparison of melting temperatures. One may then suggest to perform the calculations of the equilibrium phase diagram and the energy barriers (and the rate) of the nucleation of corresponding crystals to evaluate the GFA, as done for binary mixtures of hard disks \cite{tanaka_prx,russo_prl}. Because we have multiple parameters, $c_{\rm B}$, $p$, and $\Delta$ at fixed $\gamma$, it would be a rather difficult task.

Here we realize the comparison of the GFAs over pressures, interactions, and particle compositions in a much simpler way. The key is the finding of the underlying connection between the GFA and the $c_{\rm B}$-independent melting temperature gap between species derived from the simple conversion of units. The melting temperatures adopted here are not the actual ones from direct simulations or calculations of equilibrium systems, which should depend on not only $p$ but also $c_{\rm B}$ for given $\gamma$ and $\Delta$, but they turn out to work well to reveal the pressure and interaction dependence. Another important finding is $\Delta^*$, which suggests an energy strategy to fight against the weakening of the GFA caused by the softness of particles.

Although we do not show exact results of the equilibrium phase diagram and melting temperatures in the complicated parameter space, the schematic plots in Fig.~\ref{fig:fig5}d already to some extend indirectly reflect the pressure and interaction evolution of the equilibrium phase diagram in the $T-c_{\rm B}$ plane. It would be expected that when $p<p_{\rm e}$ the phase diagrams at various pressures look similar for $\Delta=0$ and $\Delta^*$, only that the melting temperatures grow linearly with the pressure. When $p>p_{\rm e}$, the phase diagram starts to deviate for $\Delta=0$, while the similarity may be still maintained for $\Delta=\Delta^*$.

\vspace{3mm}

\noindent {\bf Discussion}

\noindent By investigating the pressure and interaction dependence of the GFA, we find similarities and distinctions between two types of widely studied model glass-formers, binary mixtures of harmonic and RLJ particles with $\gamma=1.4$, $c_{\rm B}=0.5$, and $\Delta=0$. We focus on the energetic frustration, taking the hard particle limit as the reference. For a given $\gamma$, the GFA is the best in the hard particle limit, purely determined by geometric frustration. The involvement of the potential energy normally weakens the GFA of hard particles. We find that the GFAs of both harmonic and RLJ systems remain constant and identical at low pressures, but bifurcate at high pressures. In contrast to RLJ systems which still maintain an almost constant GFA at high pressures, significant phase separation occurs in harmonic systems.

With the variation of both $\gamma$ and $\Delta$, we come up with a generalized picture, which explains the pressure and interaction dependence of the GFA. Our major findings include i) the non-equilibrium GFA of binary mixtures of soft particles is connected to the equilibrium melting temperature, ii) the GFA of the hard particles can be maintained to high pressures with a proper modulation of the particle interaction such as using $\Delta^*$ in Eq.~(\ref{eq_deltac}), and iii) melting temperatures more linear in pressure are better to suppress phase separation and maintain a good GFA. We find that these results are valid for different values of $c_{\rm B}$. In combination of the $c_{\rm B}$ dependence studied as well in previous approaches and the pressure and interaction dependence studied in this work, we are able to have a much more comprehensive picture of the GFA of binary mixtures of soft particles. Harmonic and RLJ potentials studied here are rather different, but their behaviors can both be understood by the same picture. Moreover, the derivations in this work do not aim at any specific interaction. Therefore, we expect that our findings are valid to other types of interaction.


Note that in this work $\epsilon_{\rm AA}$, $\epsilon_{\rm BB}$, and $\epsilon_{\rm AB}$ are connected by the variable $\Delta$ and are thus not independent. Therefore, the major conclusions made here may not be directly generalized to other conditions, for example, when all three $\epsilon$'s are independence of each other. However, the robust and consistent evidence shown in this work suggests that the underlying connections between the GFA and the melting temperature should not be a coincidence. Follow-up studies are required to find out whether a more general picture can be established.

Since the melting temperature affects the GFA, it is then straightforward to expect that it also plays some role in dynamics of supercooled liquids. It is interesting to know whether $A$ and $B$ particles exhibit different dynamics related to their melting temperature difference. We are also curious about how the dynamics change when $\Delta$ varies from $0$ to $\Delta^*$ with the best GFA, from which we may reveal some underlying connections between the GFA and dynamical properties of glass-formers, such as kinetic fragility and dynamic heterogeneity.

\vspace{3mm}

\noindent {\bf Methods}

\noindent {\bf System information.} Our systems contain $N/2$ $A$ and $N/2$ $B$ particles with the same mass $m$ and a size ratio $\gamma$ defined earlier. Periodic boundary conditions are applied in all directions. We consider two types of particle interactions, harmonic:
\begin{equation}
U(r_{ij})=\frac{\epsilon_{ij}}{2}\left( 1-\frac{r_{ij}}{\sigma_{ij}}\right)^{2}\Theta\left(1-\frac{r_{ij}}{\sigma_{ij}}\right), \label{eq_pot_harm}
\end{equation}
and repulsive Lennard-Jones (RLJ):
\begin{equation}
U(r_{ij})= \frac{\epsilon_{ij}}{72}\left[ \left(\frac{\sigma_{ij}}{r_{ij}}\right)^{12} -  2\left(\frac{\sigma_{ij}}{r_{ij}}\right)^{6} +1\right] \Theta\left(1-\frac{r_{ij}}{\sigma_{ij}}\right), \label{eq_pot_rlj}
\end{equation}
where $r_{ij}$ and $\sigma_{ij}$ are the separation between particles $i$ and $j$ and sum of their radii, $\epsilon_{ij}$ is the characteristic energy scale, and $\Theta(x)$ is the Heaviside step function. We set the units of mass, energy, and length to be $m$, $\epsilon_{\rm AB}$, and $\sigma_B$, so the units of time and temperature are $\sigma_{\rm B}m^{0.5}\epsilon_{\rm AB}^{-0.5}$ and $\epsilon_{\rm AB}k_{\rm B}^{-1}$ with $k_{\rm B}$ being the Boltzmann constant.

\vspace{3mm}

\noindent {\bf Dimensionless quench rate.} In this work, we compare the GFA over a wide range of pressures from $10^{-5}$ to above $0.1$ and thus have to confront the challenge to choose a reasonable quench rate. To our knowledge, this issue has not been seriously considered, because people rarely compare the GFAs at different pressures.

It has been shown that, when glass-forming liquids are quickly quenched to $T=0$, the properties of the resultant inherent structures, e.g., potential energy and stability, depend on the parent temperature $T_{\rm p}$ prior to the quench \cite{sastry,wang,helffeich,brumer}. For glass-forming liquids, there are two characteristic temperatures, the onset temperature $T_{\rm onset}$ and the glass transition temperature $T_{\rm g}$ \cite{brumer,berthier}. When $T_{\rm p}>T_{\rm onset}$ and the liquids still exhibit the Arrhenius relaxation behavior, the potential energy of the inherent structures does not vary much with $T_{\rm p}$. When $T_{\rm g}<T_{\rm p}<T_{\rm onset}$ and the liquids are supercooled and exhibit super-Arrhenius behavior, the potential energy of the inherent structures decreases with the decrease of $T_{\rm p}$. Previous results have suggested that $T_{\rm onset}$ is around $2T_{\rm g}$ \cite{berthier}.

In our study, we start with an equilibrium liquid above $T_{\rm onset}$ and apply a quench rate $\kappa=\delta T/\Delta t$. Apparently, how long the system stays in the temperature window, $(T_{\rm g}, T_{\rm onset})$, is crucial to the structures and properties of the final solid-like states. In contrast, above $T_{\rm onset}$ or below $T_{\rm g}$, the system is either an equilibrium liquid or a glass with extremely slow structural relaxation, so how long the system stays at $T>T_{\rm onset}$ and $T<T_{\rm g}$ in the accessible time scales should not significantly affect the final solid-like state.

Previous studies have shown that $T_{\rm g}\sim p$ at low pressures for systems studied in this work with $\gamma=1.4$ and $\Delta=0$ \cite{xu}. If we use the same quench rate $\kappa$ for all pressures, the time for the system to stay in the supercooled regime, $(T_{\rm g}, T_{\rm onset})$, is $(T_{\rm onset}-T_{\rm g})/\kappa\sim p/\kappa$. Compared with high-pressure systems, the low-pressure systems almost undergo no time in the supercooled regime, which is unfair for them to explore lower-energy inherent structures. Therefore, to compensate this loss at low pressures and give systems at quite different pressures comparable chances to explore lower-energy inherent structures, we need to let $\kappa\sim p$. This leads to an updated quench rate $\kappa^*=\kappa/(p\sigma_{\rm B}^d/k_{\rm B})$, where $d$ is the dimension of space. This is actually to nondimensionlize the temperature step $\delta T$ in the expression of $\kappa$ by $p\sigma_{\rm B}^d/k_{\rm B}$.

Moreover, it has also been shown that the structural relaxation time $\tau$ of the supercooled liquids studied in this work satisfies the scaling at low pressures \cite{xu}: $\tau\sqrt{p\sigma_{\rm B}^{d-2}/m}=F(k_{\rm B}T/p\sigma_{\rm B}^d)$, where $F(x)$ is the scaling function. Therefore, in order for systems at different pressures to undergo comparable structural relaxations in the supercooled liquid regime, the quench rate is required to be further divided by $p^{1/2}$. This is actually to nondimensionlize the time duration $\Delta t$ in the expression of $\kappa$ by $(p\sigma_{\rm B}^{d-2}/m)^{-1/2}$.

Then we finally obtain a dimensionless quench rate $\tilde\kappa=\kappa^*/\sqrt{p\sigma_{\rm B}^{d-2}/m}=\frac{k_{\rm B}m^{1/2}}{\sigma_{\rm B}^{3d/2-1}}\frac{\kappa}{p^{3/2}}$. For two-dimensional systems mainly studied in this work, $\tilde\kappa=\frac{k_{\rm B}m^{1/2}}{\sigma_{\rm B}^2}\frac{\kappa}{p^{3/2}}$. By using the same $\tilde\kappa$ at different pressures, we are able to compare the GFAs of systems with the pressure varying over several orders of magnitude. This is particularly crucial to compare the low pressure regimes where both the glass transition and melting temperatures are linear in pressure. The robust scaling collapse of $\chi(\Delta T_{\rm m})$ curves strongly validates the use of $\tilde\kappa$. When $\tilde\kappa$ is fixed, $\kappa\sim p^{-3/2}$, so the computational cost dramatically increases with the decrease of pressure. As a compromise, we use larger values of $\tilde\kappa$ when focusing on the low-pressure systems in Figs.~\ref{fig:fig3}, \ref{fig:fig4}, and \ref{fig:fig5} than those for higher pressures in Fig.~\ref{fig:fig1}.

\vspace{3mm}

\noindent {\bf Data availability}

\noindent The data that support the findings of this study are available from the corresponding author upon reasonable request.

\vspace{3mm}

\noindent {\bf Code availability}

\noindent The computer codes of this study are available from the corresponding author upon reasonable request.

\vspace{1cm}

\noindent{\bf Acknowledgements}

\noindent We thank Ludovic Berthier for bringing us attention to the micro-phase separation in binary mixtures of harmonic particles at high densities, which stimulates us to perform this study. This work was supported by the National Natural Science Foundation of China Grant No.~11734014. We also thank the Supercomputing Center of University of Science and Technology of China for the computer time.

\vspace{3mm}

\noindent{\bf Author contributions}

\noindent N.X. designed the project. Y.N. performed the simulations. Y.N., J.L., J.G. and N.X. analyzed the data. Y.N. and N.X. wrote the paper.

\vspace{3mm}

\noindent{\bf Additional Information}

\noindent {\bf Competing interests:} The authors declare no Competing Financial or Non-Financial Interests.

\end{document}